# A Hybrid Trajectory Clustering
# for Predicting User Navigation

Hazarath Munaga[*1], J. V. R. Murthy[1], and N. B. Venkateswarlu[2]
[1]Dept. of CSE, JNTU Kakinada, India
Email: {hazarath.munaga, mjonnalagedda}@gmail.com
[2]Dept. of CSE, AITAM, Tekkali, India
Email: venkat_ritch@yahoo.com

*Abstract*— **In this paper, we present a novel technique for predicting and visualizing users' future navigations. Here, user navigation is considered as the sequence of URL's visited by the user. We have used distance-based specific trajectory clustering to partition users and integrated with Markov model for predicting users' future navigation. For testing the proposed technique, we developed a tool called P-NAS (Predicting and visualizing user NAvigationS), for predicting and visualizing user future navigations. We have demonstrated the effectiveness of our solution by testing the tool on user navigation data obtained from msnbc.com, and validated the result with Cross validation and Bootstrapping techniques.**

## I. INTRODUCTION

One of the greatest challenges for computer science analysts is the understanding of human behavior in the context of "digital environments" (e.g. web site). In the recent years, this knowledge is used in various means in web site development to increase ROI (return of interest) of a business organization. Predicting user navigation can be useful in many applications. For e.g. in case of web page access:

- prediction can make a change in the web advertisement (hereinafter, advt) area where a considerable amount of money is paid for placing advts' on the web sites,
- prediction helps the site designers to reorganize the web sites for enhancing the site topology and user personalization as well as semantic modeling, and
- also helpful for caching the predicted page for faster access, subsequently for improving browsing performance.

Various techniques have been discussed in the literature to cluster user sessions [1][2][3][4][5][6].

Reference [3] proposed a graph partition algorithm (Metis) that combines both the time spent on a page and Longest Common Subsequences (LCS) to cluster user sessions. The LCS algorithm has first applied on all pairs of user sessions, later each LCS path is compacted using a concept-category of page hierarchy, similarities between LCS paths were computed as a function of the time spent on the corresponding pages in the paths weighted by a certain factor. Then, it built an abstract similarity graph for the set of sessions to be clustered. Finally, Metis is used to segment the graph into clusters.

Reference [1], demonstrated the usage of hierarchical clustering algorithm (BIRCH) for clustering generalized sessions. Even though the traditional clustering techniques like k-means have been used for predicting user navigation patterns, typically, they are not very successful in attaining good results. Moreover, the k-means (i) is not suitable for generating non-globular clusters and, (ii) it is difficult to compare the quality of the computed clusters (e.g. the different initial partitions and the k value affect the outcome). Finally, the k-means approach requires lot of computational time for convergence.

Hence, some researchers attempted to improve the user navigation prediction accuracy by combining different techniques, for e.g., [2], [4], combined clustering with association rules, and [5], [6] combined clustering with Markov model.

Reference [5] partitioned site users using a model-based clustering approach where they implemented first order Markov model using the Expectation-Maximization (EM) algorithm. As per [7], EM algorithm requires a large number of iterations to take final result; hence it is very slow in convergence.

Reference [6] able to generate Significant Usage Patterns (SUP) from clusters of abstracted Web sessions; SUPs are derived from first order Markov model with each group of user sessions. In many applications, intuitively, first-order or second-order Markov models are not very accurate in predicting the user's browsing behavior, since these models do not look far into the past to correctly classify the different observed patterns. As a result, higher-order models are often used; as a result, good prediction requires higher-order models (e.g., third, fourth-order). Unfortunately, these higher-order models have a number of limitations [8] associated with high state-space complexity, reduced coverage, and sometimes even worse prediction accuracy.

Reference [9], [10] demonstrated the usage of trajectory clustering for visualizing, analyzing and obtaining hidden patterns from user navigations obtained from virtual environments and selecting cluster heads which implicitly used to extenuate the life time of wireless sensor networks respectively.

---

[*] Hazarath Munaga *alias* MHM Krishna Prasad

To overcome the above limitations, or at least a way to minimize the usage of Markov model, in this paper, we propose a specific trajectory clustering algorithm to cluster user navigation patterns of a web site and integrated with $k^{th}$ order Markov model (*hereinafter, KMM*) to predict the user navigations. Moreover, we have developed java based tool for Predicting and visualizing user NAvigationS (hereinafter, P-NAS) to predict the behavior of individual navigating in a particular web site.

## II. Specific Trajectory Clustering

The proposed algorithm can be used for predicting the user future navigations required for the analyst to answer the queries like *"what are the future visits of the user?"*.

We consider the page sequences of the site visitor as a trajectory. Any clustering algorithm requires a dissimilarity method for calculating dissimilarity between entities such as trajectories. The following section explains about the trajectory similarity employed in our algorithm.

### A. Trajectory Dissimilarity

In the literature, we have time warping distance which is used in matching speech signals in speech recognition [11]. Also, similar methods are used in DNA matching in bio-informatics. A similar technique is used to find longest common subsequence of two sequences using fast probabilistic algorithms, and then the distance is calculated [12], [13].

Here, we used Levenshtein distance [14] to calculate dissimilarity between trajectories. For creating trajectories we used simple mapping of page categories like 1 for "frontpage", 2 for "news", 3 for "tech" etc. We used Levenshtein distance to overcome the influence of large numbered pages over lower numbered ones. Table I(A) shows the sample trajectories and Table I(B) shows the obtained dissimilarity matrix for the sample trajectories with respect to *testTraj*. The dissimilarity measure which we have used in our study is given in Fig. 1 which holds metric space requirements.

TABLE I (A)
SAMPLE TRAJECTORIES

| Label | Trajectory | Consider portion for cal. dissimilarity |
|---|---|---|
| *testTraj* | 1 2 3 | 1 2 3 |
| A | 1 2 3 4 5 | 1 2 3 |
| B | 1 2 3 5 6 | 1 2 3 |
| C | 2 3 4 | 2 3 4 |
| D | 6 7 8 9 10 | 6 7 8 |
| E | 12 13 14 15 16 | 12 13 14 |

TABLE I (B)
DISSIMILARITY MATRIX FOR SAMPLE TRAJ. W.R.T TO TESTTRAJ

| Trajectory | A | B | C | D | E |
|---|---|---|---|---|---|
| *testTraj* | 0 | 0 | 3 | 3 | 3 |

*Algorithm (Compute Dissimilarity)*
*testTraj* is a trajectory having m symbols
*A* is a trajectory having n symbols; *n>=m*
Compare first *m* symbols of *testTraj* with that of *A* and find number of mismatches *(NM)*.
Dissimilarity of *testTraj* and *A* is given as:
*dis(testTraj, A)=max(NM, (m-n))*

Fig. 1 Algorithm for computing dissimilarity

### B. Probabilistic Models

The probabilistic models we have used are Markov models. These are very commonly used in the prediction of user navigations based on the previous navigations, in particular identification of the next page to be accessed by the Web site user based on the sequence of previously accessed pages [8]. For e.g., let $P = \{p_1, p_2, ..., p_m\}$ be a set of pages in a website, and $W$ be a user session with a sequence of pages visited by the user in a visit. Assuming that the user has visited $l$ pages, then $prob(p_i/W)$ is the probability that the user visits page $p_i$ next. The users next visit $(p_{l+1})$ is estimated by:

$$p_{l+1} = max_{p \in P}\{P(P_{l+1} = p/W)\}$$
$$= max_{p \in P}\{ P(P_{l+1} = p/p_l, p_{l-1}, p_{l-2}, ..., p_1)\}$$

This probability, $prob(p_i/W)$, is estimated by using all sequences of all users in training data, say $D$. Naturally, the longer $l$ and the larger $D$, the more accurate $prob(pi/W)$. However, it is infeasible to have very long $l$ and large $D$ and it leads to unnecessary complexity. Therefore, to overcome this problem, a more feasible probability is estimated by assuming that the sequence of the web pages visited by users follows a Markov process.

The fundamental assumption of predictions based on Markov models is that the next state is dependent only on the previous $k$ states. Then the above equation becomes, $P_{l+1} = max_{p \in P}\{P(P_{l+1} = p/p_l, p_{l-1}, p_{l-2}, ..., p_{l-(k-1)})\}$, where $k$ denotes the number of the preceding pages. The resulting model of this equation is called the $k^{th}$ *order Markov model* or simply KMM.

For a given sample, *the associated probability (hereinafter, AP) of the symbol is estimated by the number of times that symbol (say, $p_i$) is associated by $k$ letters to all other symbols (say, $p_1$ to n) followed by $k$ letters* i.e.

$$AP(p_i) = \frac{Frequency(p_i)}{\sum_1^n Frequency(p_i)},$$

where $n$ is no. of discrete symbols. For e.g., let us consider the sequence of $k$ letters are followed by 2 1 4 3 5 6 1 2 2 7 (10) symbols. The symbol and its associated probabilities are shown in the following Table II.

TABLE II
SAMPLE SYMBOLS AND THEIR ASSOCIATED PROBABILITY

| Symbol | 1 | 2 | 3 | 4 | 5 | 6 | 7 |
|---|---|---|---|---|---|---|---|
| AP | 0.2 | 0.3 | 0.1 | 0.1 | 0.1 | 0.1 | 0.1 |

### C. Clustering Routine

The trajectories which are having at least the length of the *testTraj* (i.e. *testTraj* contains current user navigations) will be considered for performing clustering task. The cluster routine provides the following data for the analyst to predict the navigational behavior of users:-

- A group of trajectories containing similar navigation behavior, and
- Associated probability matrix for future visits.

The clustering routine contains the following stages:

1. Dissimilarity matrix for trajectories w.r.t *testTraj* will be computed using algorithm shown in Fig. 1,
2. Using the following specific trajectory clustering algorithm, probable cluster will be computed
3. Take a trajectory (sequentially), if the dissimilarity with the *testTraj* is zero then add to the cluster C.
4. Using the following Algorithm (*Compute APMatrix*), compute associated probability of pages for future visits.
5. If there is no future probability or the obtained probability is less than the considerable value, then only build the KMM for getting future visits.

*Algorithm(Compute APMatrix)*
1. *Input:*
   $C$ is a cluster having $n$ trajectories;
   *testTraj* is a trajectory containing current user visits;
2. *Output:*
   $AP$ is an associated probability, a one-dimensional matrix of size ($m$) for $m$ symbols;
   *next* is a one-dimensional matrix of size $m$ used for counting next/immediate probable page weights;
   *temp = 0, totalCount = 0* are temporary variables;

   //for getting the immediate page id
3. for $i = 1$ to $n$ of $C$,
   *temp = C[i][testTraj.length+1];*
   *next[temp]++;*
   *totalCount++;*
   *end for*

   // for computing the AP of pages
4. for $i = 1$ to $m$ symbols
   *AP[i] = next[i]/totalCount;*
   *end for*
5. return *AP*;

### III. EXPERIMENTAL WORK

We have used the Web navigation data obtained from "msnbc.com"[2]. The data comes from Internet Information Server (IIS) logs and news related portions of msnbc.com for the entire day of September, 28, 1999 (Pacific Standard Time). Each sequence in the dataset corresponds to page views of a user during that twenty-four hour period. Each instant in the sequence corresponds to a users request for a page. Requests are recorded at the level of page category.. The categories in "frontpage", "news", "tech", "local", "opinion", "on-air", "misc", "weather", "health", "living", "business", "sports", "summary", "bbs" (bulletin board service), "travel", "msn-news", and "msn-sports". The full dataset consists of 989818 (users), with an average of 5.7 events per sequence. The computed detailed view of the dataset (no. of pages visited by user') is shown in following Table Fig.2. In our study, we have taken the potential users who visited at least 3 pages and up to 13 pages; it got up to 173879 users.

| No. of Visits | Users | % with total | No. of Visits | Users | % with total |
|---|---|---|---|---|---|
| 1 | 601384 | 60.76 | 10 | 297 | 0.03 |
| 2 | 214392 | 21.66 | 11 | 142 | 0.01 |
| 3 | 94711 | 9.57 | 12 | 238 | 0.02 |
| 4 | 43321 | 4.38 | 13 | 143 | 0.01 |
| 5 | 19692 | 1.99 | 14 | 74 | 0.01 |
| 6 | 8902 | 0.90 | 15 | 67 | 0.01 |
| 7 | 4008 | 0.40 | 16 | 9 | 0.00 |
| 8 | 1688 | 0.17 | 17 | 13 | 0.00 |
| 9 | 737 | 0.07 | | | |

Fig. 2 Detailed view of the user visits

For testing our proposal, as a test case, we consider already visited and currently visited pages are 1, 3 and 4; the following is obtained output from the tool:

- Fig. 3a., shows the extracted clusters from the dataset,
- As shown in Fig. 3b, c, after visiting 1 3 4 pages, the user is expected to visit $2^{nd}$ page with 57%, $7^{th}$ page with 26% and $12^{th}$ page with 17% probability,
- If the user visits the $2^{nd}$ page; as shown in Fig. 3d, he can further expected to visit $12^{th}$ page with 50%, $6^{th}$ page with 29% and $14^{th}$ page with 21% probability

*Even though our tool is designed to predict future user navigations, it supports the administrator/ website designer in the following way:*

1. Based on the status of the user, in advance next probable pages can be brought to the cache, subsequently it reduces the cache latency and access time. Moreover, if the required page is available in cache /main memory latency is very small, on the other hand it should be brought from secondary memory leading to page faults. If we predict properly these types of problems can be avoided or at least reduced.
2. The advts' or the things which require maximum user attention can be dynamically positioned in the next probable pages.
3. If a potential company wants to give advt in more than one page, the tool suggests most probable pages:
   - for e.g., as shown in Fig. 3b, after visiting 1 3 4 pages, users are going to navigate through 2nd, $7^{th}$ and 12th pages; hence, the company can present its advt not only in 4th page, and it can be repeated or presented in more detail way in 2nd, 7th and $12^{th}$ pages to get more users attention,

ACEEE

- on the other hand, if the company can't invest the huge amount (in general, cost for advt in "main" or "business" page is expensive when compare with other pages) still it can present its advt. For e.g., we observe that, users visited "news" and "business" pages, with 40% probability they are going to visit the "tech" page. The company can present its advt as a small snapshot in "business" page later he can present the detail advt in the next probable page i.e. "tech" page.

### A. Validation

Since the 1980s, cross-validation (CV) [16] and bootstrapping (BTS) [17] have been the popular techniques for estimation of the unknown performance of a classifier designed for discrimination. Here we used the 5-fold CV and BTS to validate our model. Obtained results from the 5-fold validation are shown in Fig. 4, where it shows the achieved percentage of success and average clusters formed for corresponding user sessions. The proposed model is tested for a variety of contexts, and observes that:

1. As shown in Fig. 4a, for a mixed dataset of users who navigated between 3 to 8 pages for predicting future visit with KMM support we achieve a maximum of 95% success. In particular, our goal is to reduce the usage of KMM model, due to its huge requirement of memory and processing time. For predicting the users 4th visits almost we achieve same percentage of success without using KMM.

2. As shown in Fig. 4b., for predicting users exact visit between 4th to 10th, for predicting 4th visit

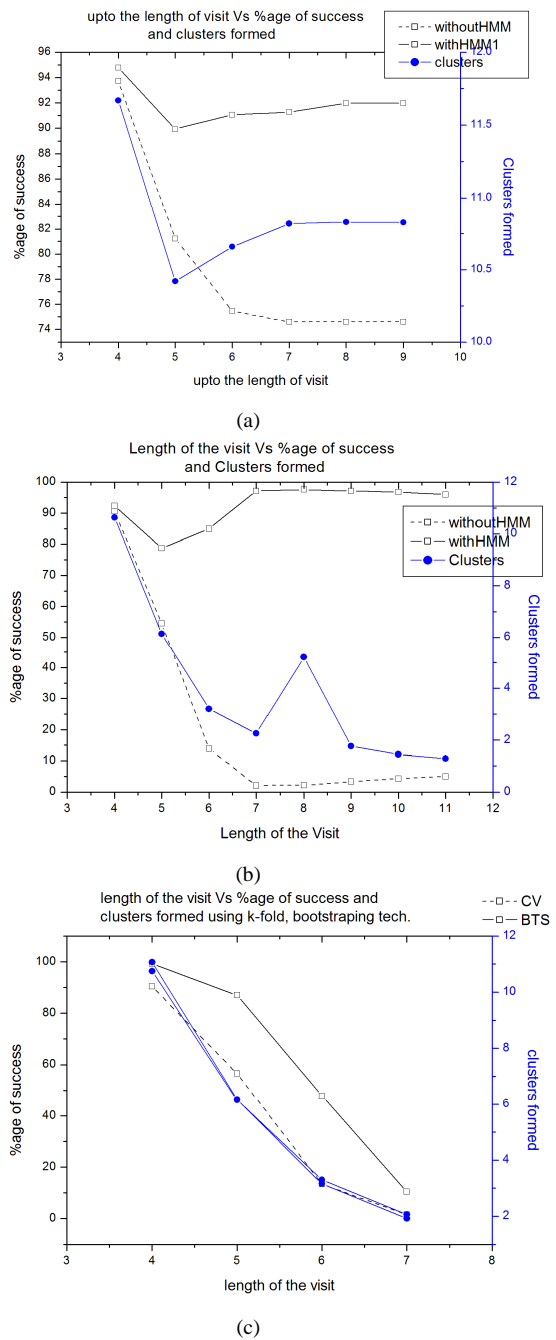

(a)

(b)

(c)

Fig. 4. (a) Validation Results for mixed dataset (b) Validation Results for particular visit (c) Comparison between BTS and 5-fold CV techniques

without KMM support we got 94% success; and as usual, as the length of visit increases success rate is drop down, and in this context, it becomes mandatory to use KMM model for achieving good success rate.

3. To verify, the obtained validation result using 5-fold CV technique, for the same dataset without using KMM support, we compare with BTS technique; from Fig. 4c, we can easily observe that always BTS is giving higher success rate and almost similar clusters for the dataset.

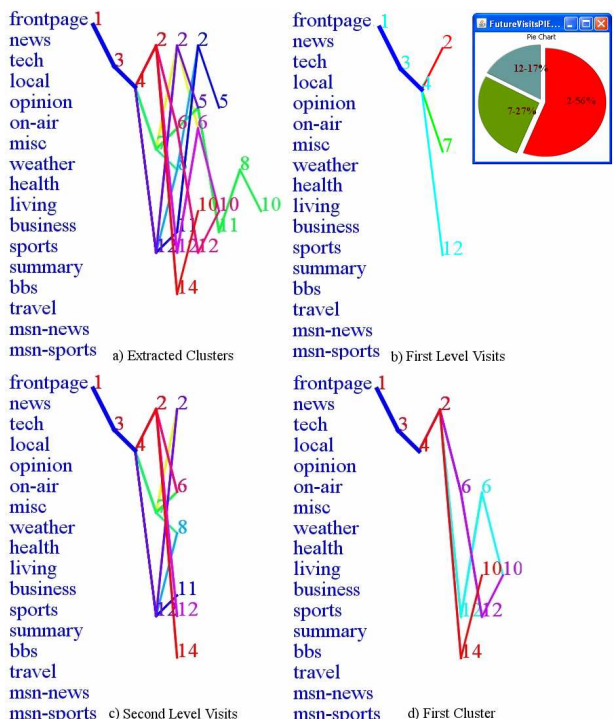

Fig. 3 Obtained visualization from the P-NAS

## IV. CONCLUSION

In this paper, we have presented a novel specific trajectory clustering algorithm for clustering, visualizing and analyzing user future navigation patterns. Users are grouped into clusters such that the users with similar navigation patterns are placed in the same cluster. Experiments are carried out using the msnbc.com users' data, and validated with cross-validation (CV) and bootstrapping (BTS) techniques. This tool can be used for predicting the user future visits. This knowledge can be used in increasing the business.

This model can be modified to make further improvement in prediction accuracy. An important observation that I made while going though the literature on various web prediction models is that, not many researchers have made an attempt to optimize the model using genetic algorithms or simulated annealing, hence, it can be considered as a future work. This model has a drawback, i.e., it does not track the probability of the next item that has never been seen. Using a variant of prediction by partial matching will help take care of this situation and should be considered in the work ahead.

## REFERENCES

[1] Y. Fu, K. Sandhu, and M.-Y. Shih, "Clustering of web users based on access patterns," in *Proceedings of the 1999 KDD Workshop on Web Mining*. Springer-Verlag, 1999.

[2] H. Lai and T.-C. Yang, "A group-based inference approach to customized marketing on the web integrating clustering and association rules techniques," in *HICSS '00: Proceedings of the 33rd Hawaii International Conference on System Sciences*-Volume 6. Washington, DC, USA: IEEE Computer Society, 2000, p. 6054.

[3] A. Banerjee and J. Ghosh, "Clickstream clustering using weighted longest common subsequences," in *Proceedings of the Web Mining Workshop at the 1st SIAM Conference on Data Mining*, 2001, pp. 33–40.

[4] S. L. F Liu, Z Lu, "Mining association rules using clustering," *Intelligent Data Analysis*, vol. 5, no. 4, pp. 309–326, 2000.

[5] I. Cadez, D. Heckerman, C. Meek, P. Smyth, and S. White, "Modelbased clustering and visualization of navigation patterns on a web site," *Data Min. Knowl. Discov.*, vol. 7, no. 4, pp. 399–424, 2003.

[6] L. M. Dunham, and Y. Meng, "Discovery of significant usage patterns from clusters of click - stream data," in *WebKDD2005*, 2005.

[7] L. Xu and M. I. Jordan, "On convergence properties of the em algorithm for gaussian mixtures," *Neural Comput.*, vol. 8, no. 1, pp. 129–151, 1996.

[8] M. Deshpande and G. Karypis, "Selective markov models for predicting web page accesses," *ACM Trans. Interet Technol.*, vol. 4, no. 2, pp. 163–184, 2004.

[9] H. Munaga, L. Ieronutti, and L. Chittaro, "CAST - a novel trajectory clustering and visualization tool for spatio temporal data," in *IHCI-2009: Proceedings of the First International conference on Intelligent Human Computer Interaction*. Springer, India, January 2009, pp. 169–175.

[10] H. Munaga, J. V. R. Murthy, and N. B. Venkateswarlu, "A novel trajectory clustering technique for selecting cluster heads in wireless sensor networks," *International Journal on Recent Trends in Engineering*, vol. 1, pp. 357–361, May 2009.

[11] H. Sakoe and S. Chiba, *Dynamic programming algorithm optimization for spoken word recognition*. San Francisco, CA, USA: Morgan Kaufmann Publishers Inc., 1990.

[12] G. Das, D. Gunopulos, and H. Mannila, "Finding similar time series," in *PKDD '97: Proceedings of the First European Symposium on Principles of Data Mining and Knowledge Discovery*. London, UK: Springer-Verlag, 1997, pp. 88–100.

[13] B. Bollob´as, G. Das, D. Gunopulos, and H. Mannila, "Time-series similarity problems and well-separated geometric sets," *Nordic J. of Computing*, vol. 8, no. 4, pp. 409–423, 2001.

[14] L. I. Vladimir, "Binary codes capable of correcting deletions, insertions and reversals," *Soviet Physics Doklady*, vol. 10, pp. 707–710, February 1966.

[15] F. Khalil, J. Li, and H. Wang, "Integrating recommendation models for improved web page prediction accuracy," in *ACSC '08: Proceedings of the thirty-first Australasian conference on Computer science*. Darlinghurst, Australia, Australia: Australian Computer Society, Inc., 2008, pp. 91–100.

[16] M. Stone, "Cross-validatory choice and assessment of statistical predictions," *Roy. Statist. Soc. Ser. B*, vol. 36, pp. 111–148, 1974.

[17] B. Efron and R. Tibshirani, *Introduction to the Bootstrap*. Chapman and Hall, London, 1993.